\documentclass[%
 reprint,
 amsmath,amssymb,
 aps,
]{revtex4-2}

\usepackage[english]{babel}
\usepackage{graphicx}
\usepackage{dcolumn}
\usepackage{bm}
\usepackage{quantikz}
\usepackage{multirow}
\usepackage{subcaption}
\usepackage{braket}
\usepackage{soul}
\usepackage{hyperref}

\newtheorem{definition}{Definition}

\usepackage{tabularx}

\graphicspath{{Introduction/figures/}{QuantumUtility/figures}{PracticalAssessment/figures}{Conclusion/figures}}

\begin{document}

\title{Quantum utility -- definition and assessment of a \textit{practical} quantum advantage} 

\author{Nils Herrmann}
 \email{nils.herrmann@quantum-brilliance.com}
\author{Daanish Arya}
\author{Florian Preis}
 \email{f.preis@quantum-brilliance.com}
\author{Stefan Prestel}
\affiliation{Quantum Brilliance GmbH \\ Colorado Tower Industriestr.\ 4 \\ 70565 Stuttgart, Germany}
\author{Marcus W.\ Doherty}
\author{Angus Mingare}
\author{Jason C.\ Pillay}
\affiliation{Quantum Brilliance Pty Ltd \\ 60 Mills Road \\ Acton ACT 2601, Australia}

\date{\today}

\begin{abstract}
  Several benchmarks have been proposed to holistically measure quantum computing performance. While some have
  focused on the end user's perspective (e.g., in application-oriented benchmarks), the real industrial value 
  taking into account the physical footprint of the quantum processor are not discussed. Different use-cases come
  with different requirements for size, weight, power consumption, or data privacy while demanding to surpass 
  certain thresholds of fidelity, speed, problem size, or precision. 
  This paper aims to incorporate these characteristics into a concept coined \textit{quantum utility}, which 
  demonstrates the effectiveness and practicality of quantum computers for various applications where quantum 
  advantage -- defined as either being faster, more accurate, or demanding less energy -- is achieved over a 
  classical machine of similar size, weight, and cost. To successively pursue quantum utility, a level-based 
  classification scheme -- constituted as \textit{application readiness levels} (ARLs) -- as well as extended 
  classification labels are introduced. These are demonstratively applied to different quantum applications 
  from the fields of quantum chemistry, quantum simulation, quantum machine learning, and data analysis 
  followed by a brief discussion. 
\end{abstract}
\maketitle

\section{Introduction}
In recent years, there have been claims of quantum computers reaching quantum advantage \cite{Arute2019, Zhong2020, Zhong2021, Madsen:2022uqz} -- the point at which quantum computers are able to solve a problem that a classical device would not be able to in a reasonable amount of time \cite{Arute2019}. This is a critical turning point in the computational sciences as it precedes an era of widespread use of quantum computers.
All current quantum advantage claims reaching super-polynomial speed-ups \cite{Preskill:2012tg}, rely on either random circuit sampling \cite{10.5555/3135595.3135617} or Gaussian boson sampling methods \cite{10.1145/1993636.1993682}. 
These can be viewed as extreme niche use-cases with questionable practical relevance.
Nevertheless, they demonstrate that there is no fundamental physical law which prevents quantum advantage from being achieved. 

In the past endeavors to establish a genuine quantum advantage, the physical demands of quantum processors in terms of size, weight, and power consumption and thus
costs (SWaP-C) are rarely considered if at all. Only if these SWaP-C factors are taken into consideration, users are enabled to decide which computing platform is suitable and beneficial for a particular application. More community focus on SWaP-C could thus lead to the widespread adoption of quantum computing. Therefore, we introduce the idea of a \textit{practical} quantum advantage -- called quantum utility --, which 
incorporates device specifications in terms of SWaP-C demands.

In terms of general performance benchmarks for quantum computing, one of the first metrics -- called quantum volume\cite{2018QS&T....3c0503M, Cross2019} -- utilized by multiple quantum computing companies was introduced by IBM.
It considers various properties of the device, such as number of qubits, circuit compilation, coherent and non-coherent errors, gate parallelism, device connectivity and computational efficiency to create a single-number value to quantify the abilities of a quantum computer.
In addition to quantum volume to measure the quality and scale of a quantum computer, the concept of circuit layer operations per second (CLOPS) has been subsequently introduced \cite{wack2021}.
CLOPS builds on the quantum volume circuits and determines the number of circuit layers which can be executed per second. 

Whilst quantum volume can provide information on the performance of the machine, its deployment is limited to the performance of only square circuits, which represents a compromise of the large variety of algorithms of interest such as the quantum Fourier transform, quantum phase estimation, Deutsch-Jozsa's, Grover's, and Shor's algorithm \cite{nielsen00} or many variational quantum algorithms \cite{Cerezo:2020jpv}. These are realized as either shallow or deep rectangular circuits in general.
Consequently, the notion of the quantum volume proposal was generalized to so-called volumetric benchmarks \cite{BlumeKohout2020volumetricframework, Mills2021application}.

It is important to note that these benchmarks rely on the same theoretical framework for random circuit sampling \cite{10.5555/3135595.3135617} as references \cite{2018NatPh..14..595B, Arute2019} for demonstrating quantum advantage. 
As noted in references \cite{PhysRevLett.117.170502,Proctor2022}, randomized circuits are less sensitive to coherent errors, which are amplified in structured circuits. 
While this fact can be exploited in randomized compilation \cite{PhysRevA.94.052325} it is not desired in benchmarking, since many quantum algorithms such as the variational quantum eigensolver \cite{peruzzo2014variational} with many repeating blocks are instances of structured quantum programs.

While random circuits can mimic some aspects of variational circuits, the potential for applications of random circuit sampling is as of yet unclear. Therefore, several proposals towards application-oriented benchmarking have been made.

Q-Score was proposed as a method of defining the maximum number of qubits a quantum computer has available to run a max-cut algorithm. This demonstrates the size of problem that a processor is able to solve \cite{Martiel2021}.

The benchmarking suite introduced in reference \cite{Lubinski2021} contains the three categories: tutorial (e.g., Deutsch-Jozsa), subroutine (e.g., quantum Fourier transform), and functional (e.g., variational quantum eigensolver). 
Here, the performance is measured by the averaged result fidelity and quantum execution times for each circuit.
The resulting metrics are displayed in the spirit of volumetric benchmarks (width and depth of the circuits) and are compared to a volumetric background.
Since the benchmarking suite does not contain a full-blown application, e.g., electronic structure simulations of molecules, the measured execution times are only of limited value for estimating application runtimes. 
Even the functional benchmarks do not capture the full hybrid nature of near term applications, which require many passes between classical and quantum computing resources in many iterations, optimization of circuits to specific hardware topologies, and error mitigation.

Recently, a benchmarking suite was proposed\cite{Finzgar:2022mya} focussing on end-to-end application performance and so far contains a robot path planning and a vehicle option planning problem.

A common problem of all previously discussed benchmarks is the difficulty to verify success at scale, as the comparison with a classical computation becomes increasingly difficult with an increasing width of the quantum circuits. 
In special cases such as Shor's algorithm, one can efficiently verify the result of prime factorization by multiplication. 
With the use of so-called mirror circuits\cite{Proctor2022}, this may be generalized to more common circuits. 
Here, a given circuit $C$ is appended with a layer of Pauli gates followed by the quasi-inverse  $\tilde{C}^{-1}$ of the given circuit, which should reverse the action of $C$ up to the latter Pauli operations. 

In the following section, we propose a framework to systematically assess the maturity and commercial value of quantum enhanced applications and the associated quantum algorithms that allows to drive adoption forward.
In section \ref{sec:QuantumUtility}, we introduce the notion of quantum utility and define Application Readiness Levels (ARLs) -- a corresponding measurement system.
Furthermore, we propose extensions of performance benchmarks and discuss criteria for progressing through the ARLs during the development of quantum applications.
In section \ref{sec:practicalassessment}, we apply the quantum utility framework to selected quantum applications from the fields of quantum chemistry, quantum simulation, quantum machine learning (QML), and data analysis.

\section{\label{sec:QuantumUtility}A practical quantum advantage}
Quantum computing is often praised as a breakthrough technology that will have tremendous impacts on the computational sciences (see, e.g., \cite{Doherty2021, Gill2022, Rietsche2022}). In particular, future quantum devices may enable computations deemed impossible on any available classical device -- often referred to as \textit{quantum supremacy}~\cite{Arute2019, 10.5555/3135595.3135617, 2018NatPh..14..595B} or more recently just as \textit{quantum advantage}~\cite{Zhong2020, doi:10.1126/sciadv.abl9236, Wu2021}. It is reasonable to assume that such an advantage will not appear instantaneously, but rather emerge slowly over time. A natural milestone on this path is the advent of commercially-available quantum devices that, when performing practical applications, can outperform a classical device in terms of speed, accuracy and energy consumption.

In this section, we build upon that philosophy of a practical quantum advantage. In subsection \ref{sec:QuantumUtility:Definitions}, we strictly define (and separate) the terms \textit{quantum utility} and \textit{quantum dominance}. Subsection \ref{sec:QuantumUtility:ARL} further introduces a simple concept -- \textit{application readiness levels} (ARLs) -- to characterize and assess quantum applications in terms of their utility.

\subsection{\label{sec:QuantumUtility:Definitions}Definitions}

\begin{figure*}
  \centering
    \includegraphics{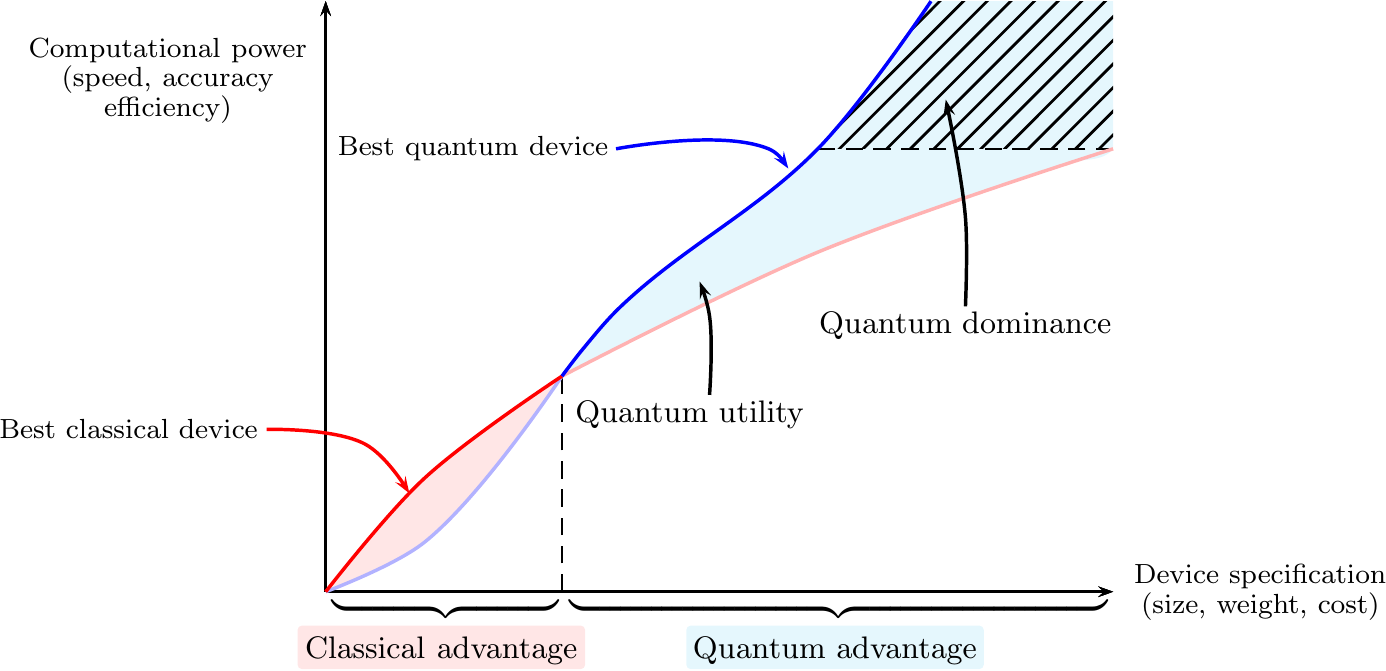}
    \caption{\label{fig:QuantumUtility}Abstract visualization of computational power (in terms of either speed, accuracy or power efficiency) w.r.t.\ device specifications (in terms of size, weight and cost) for the best classical device (red line) and best quantum device (blue line), in analogy to reference~\cite{Nguyen2022}. In contrast to the light red area of classical advantage, the light blue area of quantum advantage sub-categorizes into quantum utility and quantum dominance, depending on the availability of larger (better performing) classical devices compared to the best quantum device.}
    \end{figure*}

In this work, we introduce and reintroduce the terms \textit{quantum utility}, \textit{quantum dominance}, and \textit{quantum advantage} according to definitions \ref{def:QU}, \ref{def:QD}, and \ref{def:QA}, respectively:

\begin{definition}
\label{def:QU}
\textbf{Quantum utility} is obtained if a practical application
\begin{itemize}
    \item[(i)] requires less computing time, or
    \item[(ii)] requires less power, or
    \item[(iii)] yields more accurate results
\end{itemize}
on a quantum device or a hybrid classical/quantum architecture, compared to the \textbf{best} classical device of \textbf{similar size, weight, and cost}.
\end{definition}

\begin{definition}
\label{def:QD}
\textbf{Quantum dominance} is obtained if a practical application fulfills requirements (i)--(iii) from Definition \ref{def:QU} on a quantum device or a hybrid classical/quantum architecture, compared to \textbf{any} other classical device.
\end{definition}

\begin{definition}
\label{def:QA}
\textbf{Quantum advantage} is defined as a collective term meaning either \textbf{quantum utility} or \textbf{quantum dominance}.
\end{definition}

Fig.~\ref{fig:QuantumUtility} illustrates these definitions in terms of abstract computational power versus abstract physical device specification for the best performing classical and quantum devices. Device advantage, indicated by the curves, separate into two regions of classical and quantum device advantage. The region of quantum device advantage is further separated into a region of quantum utility and quantum dominance according to definitions \ref{def:QU} and \ref{def:QD}. In particular, quantum utility is reached whenever a quantum device outperforms a classical competitor of the \textit{same} device specifications.

While quantum dominance remains the main goal of most quantum computing companies, quantum utility may be a more realistic target for the intermediate-scale devices currently being developed. Also, quantum dominance may not be required at all to reach a genuine advantage -- especially for applications that have strict SWaP-C constraints (e.g., mobile or edge-based). Even if quantum computing will never achieve quantum dominance, quantum utility -- as long as SWaP-C constraints are met -- might still lead to an improvement of tremendous value. 
Hence, quantum utility marks a milestone of great importance for both manufacturers of quantum technology and their customers alike, at which a quantum application first takes on commercial significance.

\subsection{\label{sec:QuantumUtility:performancemetric}Performance benchmarks}

Inspired by The Green Index (TGI) \cite{Subramaniam2012} we propose to adopt similar measures of energy efficiency for quantum computing which allows users to compare different quantum computing architectures. It was seen as unsustainable to consider performance based on only computational power. The TGI allows for corporations to make environmentally considered decisions about the types of HPC systems used. The TGI employs a performance per Watt metric based on the floating point operations per second (FLOPS) per Watt. 
The FLOPS are typically measured by the performance of the LINPACK benchmark.
So far, no quantum benchmarks have taken the environmental aspect into account, but ideas around sustainable quantum computing begin to gain traction, see e.g., \cite{PRXQuantum.3.020101, Fellous2022} and the Quantum Energy Initiative.

More generally, the performance can be given by any benchmarking score, such as circuit fidelities or successful number of quantum gate operations, divided by the total runtime. Here, the runtime may be multiplied by the average power consumption of the device yielding the total energy consumed in the denominator of the benchmark, therefore, penalizing power hungry devices:
\begin{equation}
\mathrm{Score}_1 = \frac{\mathrm{Performance}}{\mathrm{Runtime} \times \mathrm{Power}}
\end{equation}

In many deployment scenarios in automotive or aerospace industries, size and weight are further essential measures.
A like-for-like comparison of devices would allow stakeholders in industries such as automotive to decide whether to replace their existing processors with a quantum equivalent.
In classical computing, Sun Microsystems introduced the Space, Wattage, and Performance metric incorporating an additional factor of volume in the denominator of their benchmark to account for the size of the computing device:
\begin{equation}
\mathrm{Score}_2 = \frac{\mathrm{Performance}}{\mathrm{Volume}\times \mathrm{Runtime} \times \mathrm{Power}}
\end{equation}

In order to assess Quantum Utility as introduced in the previous subsection we have to incorporate all dimensions into future benchmarks: quantity, quality, speed, usefulness and physical, economical as well as ecological footprint.
We recommend that the quantum computing industry adopts energy efficiency standards established in the classical computing industry. Hardware providers should disclose the power consumption, volume, and weight of the quantum computer (including control and all infrastructure required for maintaining the coherence of the qubits) in addition to fidelities, number of qubit, T1, and T2 times.

\subsection{\label{sec:QuantumUtility:ARL}Application readiness levels and their assessment}

\begin{figure*}
  \centering
\includegraphics[width=\textwidth]{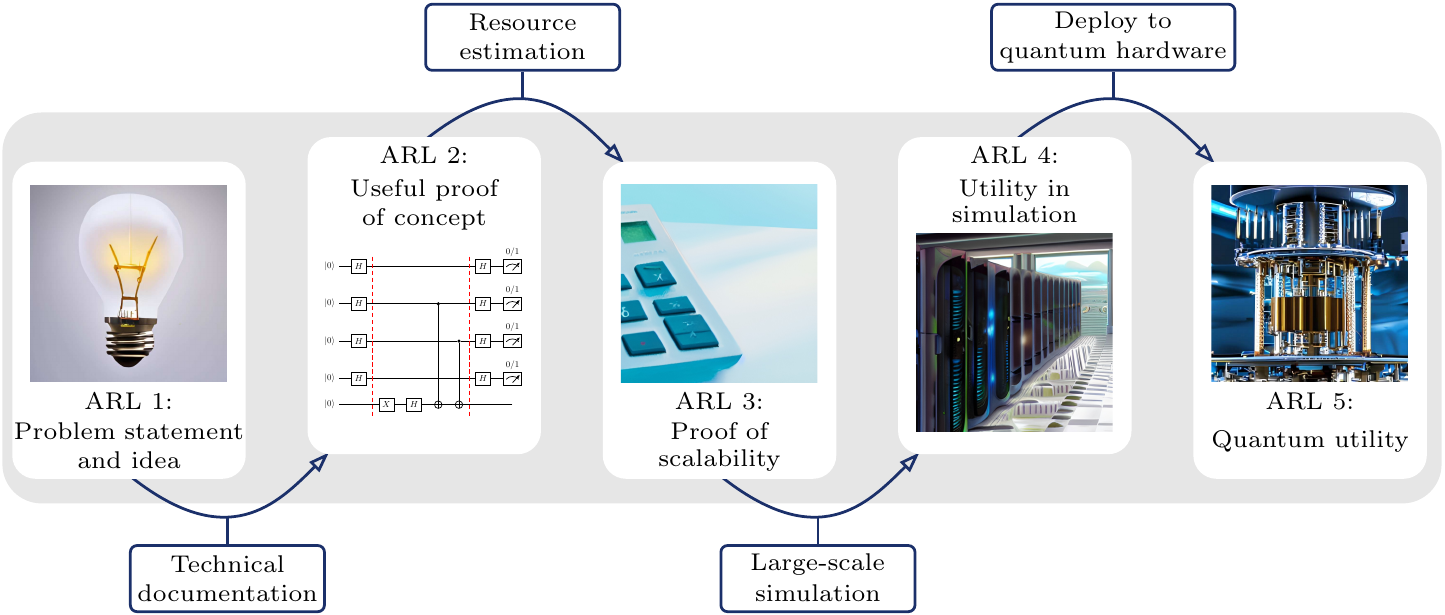}
\caption{\label{fig:ARL_Roadmap}Schematic roadmap detailing the progression through and exit strategies of the individual application readiness levels (cf. Figure \ref{fig:ARL_Milestones}) in the pursuit of quantum utility (see Def.\ \ref{def:QU}).}
\includegraphics[scale=1]{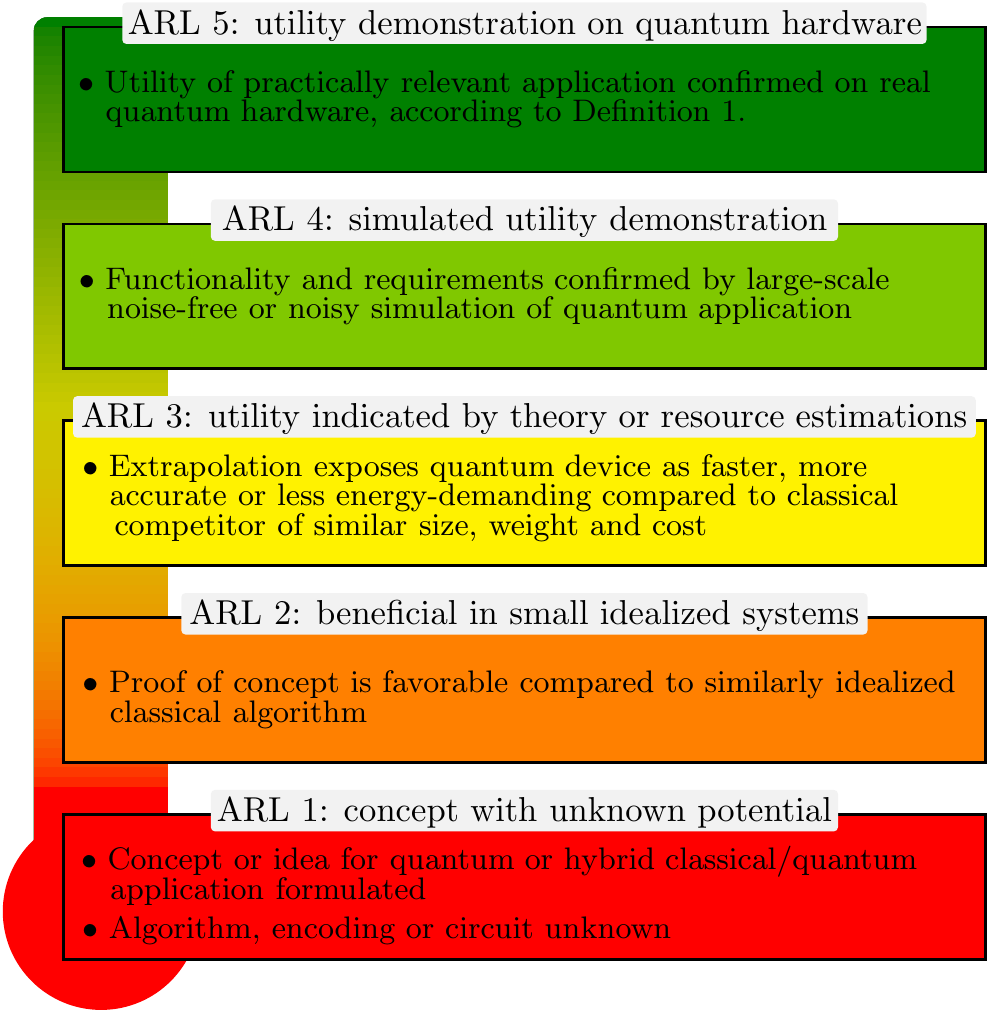}
\caption{\label{fig:ARL_Milestones}Milestones and definition of application readiness levels (ARLs) for the classification and assessment of the maturity of quantum applications in the pursuit of quantum utility (referred to as just 'utility', see Def.\ \ref{def:QU}).}
\end{figure*}

Following NASA's development of technology readiness levels (TRLs) in the late 80s, an international standard was published in 2013\cite{ISO_TRL_2013}. Ever since, TRLs are a widely accepted tool to track and assess the maturity of emerging technologies. In analogy to TRLs, we propose a five-level scheme we denote \textit{application readiness levels} (ARLs) to track the maturity of quantum applications in the pursuit of quantum utility (cf.\ Def.\ \ref{def:QU}). In Figures \ref{fig:ARL_Roadmap} and \ref{fig:ARL_Milestones}, a schematic roadmap including exit strategies through the individual ARLs as well as milestones, and definitions are highlighted, respectively.

Any idea or concept of a quantum application is categorized as ARL--1. At this level, no formal description (e.g., in terms of a quantum circuit) of that algorithm is necessary. ARL--1 is meant as an entry label to pave the way of further investigations and developments, even if the application is merely a thought. With more in-depth research and a working proof of concept implementation, the application may proceed to ARL--2 if it can present any kind of benefit compared to a (possibly scaled-down) similar-sized classical competitive algorithm. At this stage, the application may enter an advanced analysis phase, where its formal resource scaling w.r.t.\ problem-specific variables is investigated. In our own studies (cf.\ section \ref{sec:practicalassessment}) we found that an extended classification scheme in terms of a quantum application's core features may be useful. The labels of a such an extended classification scheme, which we will use in section \ref{sec:practicalassessment}, are summarized in Table \ref{extended_labels.tab}.
\begin{table}
  \centering
\caption{\label{extended_labels.tab}Additional classification labels for the analysis of ARL--2 applications.}
\begin{tabular}{c|l}
\textbf{Category} & \textbf{Labels} \\
\hline
\multirow{4}{*}{Scalability} & $\mathcal{O}$ scaling of \\
& (i) the number of measurement circuits, \\
& (ii) the gate depth per circuit, and \\
& (iii) the number of shots per circuit \\ \hline
\multirow{5}{*}{Compilability} & $\bullet$ Native gates \\
& $\bullet$ Non-native 1 and/or 2 qubit gates \\ 
& $\bullet$ General multi-qubit gates \\ 
& $\bullet$ Classical control \\ \hline
\multirow{4}{*}{Connectivity} & $\bullet$ Linear \\ 
& $\bullet$ Circular \\ 
& $\bullet$ Nearest neighbor \\ 
& $\bullet$ All-to-all \\ \hline
\multirow{3}{*}{Robustness} & $\bullet$ Noise as a resource \\ 
& $\bullet$ Variational circuits \\ 
& $\bullet$ Non-variational circuits \\ \hline
\multirow{3}{*}{Parallelizability} & $\bullet$ Qubit-based \\ 
& $\bullet$ Circuit-based \\ 
& $\bullet$ Shot-based \\ 
\end{tabular}
\end{table}
These include 
\begin{itemize}
    \item Scalability: Given a set of problem-specific variables (e.g., the number of required qubits), it is important to understand how demanding a quantum application becomes with increasing problem sizes. An extensive scaling study should therefore involve the formal scaling of (i) the total number of quantum circuits that need to be executed, (ii) the number of successive quantum gate operations, i.e., the gate depth, to be executed in each circuit, and (iii) the number of measurement shots required to sample the quantum state with sufficient accuracy. 
    \item Compilability: When working with real quantum hardware, one has to compile the algorithm gates into the available native hardware gates. Depending on the original algorithm gates (and the available native gates) this may be a more or less demanding task. In general one may distinguish between an algorithm that uses only native gates, uses non-native 1- and 2- qubit gates, requires general multi-qubit gates, or involves some degree of classical control such as conditional mid-circuit measurements or classical oracles.
    \item Connectivity: A given quantum hardware may only be able to entangle specific pairs or sets of qubits. To make optimal use of this constrained qubit topology, the algorithmic qubits need to be mapped in an optimal fashion to the real physical qubits. Depending on the connectivity requirements of a given quantum circuit, this may lead to significant increases in compiled gate depths. In general, one may distinguish between algorithms that require linear, circular, nearest neighbor (in 2 or 3-dimensional space), or an all-to-all connectivity. 
    \item Robustness: All currently developed quantum computers are subject to noise. The robustness of a given quantum application to this noise may be of great importance. In general, one may distinguish between applications that use noise as a resource, involve trained parametrized circuits that may to some degree adapt to the presence of noise, and general non-variational circuits that assume perfect or near perfect execution. 
    \item Parallelizability: Due to the probabilistic nature of quantum computing, any quantum application is inherently parallelizable. The degree of parallelization, however, depends on the underlying quantum circuits. In general, one may distinguish between a qubit-based parallelizability, where the quantum circuits may be split into smaller circuits requiring less qubits, a circuit-based parallelizability, where a specific application involves the simultaneous measurement of multiple circuits, and a pure shot-based parallelizability where a given quantum application may be deployed to multiple quantum processing units to increase only the obtained shot rate. 
\end{itemize}

Once an application is properly assessed according to these extended classification labels, the resource requirements for specific quantum hardware may be extrapolated. If such an extrapolation reveals an advantage over classical competitors, the application reaches ARL--3. From here on, one may prove these theoretical extrapolations by performing simulations (if possible) or actual quantum computations to reach ARL--4 and ARL--5, respectively. ARL--4 is further subcategorized into noise-free (ARL--4a) and noisy (ARL--4b) simulations. The final target, however, -- quantum utility -- is reached only if an application reaches ARL--5. 

\section{\label{sec:practicalassessment}Practical Assessment}

\begin{table*}
  \centering
    \caption{\label{tab:QUSurvey}Classification of selected quantum applications according to their application readiness level (ARL, cf.\ Figures \ref{fig:ARL_Roadmap} and \ref{fig:ARL_Milestones}) as well as their extended classification labels (cf.\ Table \ref{extended_labels.tab}). If multiple circuits / algorithms are available, only the optimal extended labels are shown.}
    \def\arraystretch{1.5}
    \begin{tabular}{c|c|ccc|c|c|c|c}
        \textbf{Application} & \textbf{ARL} & \multicolumn{3}{c|}{\textbf{Scalability}} & \textbf{Compilability} & \textbf{Connectivity} & \textbf{Robustness} & \textbf{Parallelizability} \\
        & & \textbf{\#Circuits} & \textbf{Depth} & \textbf{\#Shots} &&&& \\
        \hline 
        \multicolumn{9}{c}{\rule{0pt}{11pt}Quantum chemistry and quantum simulation} \\[3pt]
        \hline
        VQE\cite{peruzzo2014variational} & 3 & $\mathcal{O}(N)$ & $\mathcal{O}(N)$ & $\mathcal{O}(1)$ & native gates & linear & variational & circuit-based \\ \hline
        QRBM\cite{Xia2018} & 2 & $\mathcal{O}(1)$ & $\mathcal{O}(nm)$ & $\mathcal{O}\binom{n}{n_p}$ & classical control & all-to-all & variational & shot-based \\ \hline
        VarQiTE\cite{McArdle2019} & 2 & $\mathcal{O}(tq(q+p))$ & $\mathcal{O}(q)$ & $\mathcal{O}(1)$ & non-native gates & all-to-all & variational & circuit-based \\ \hline 
        \multicolumn{9}{c}{\rule{0pt}{11pt}Binary and multi-class classification} \\[3pt] 
        \hline
        QK\cite{Schuld2019} & 2 & $\mathcal{O}\binom{|T|}{2}$ & $\mathcal{O}(N)$ & $\mathcal{O}(2^N)$ & non-native gates & linear & non-variational & circuit-based \\ \hline 
        QVC\cite{Schuld2019, Havlicek2019} & 2 & $\mathcal{O}(|T|)$ & $\mathcal{O}(N)$ & $\mathcal{O}(1)$ & non-native gates & linear & variational & circuit-based \\ \hline 
        Re-Uploading\cite{PerezSalinas2020} & 2 & $\mathcal{O}(|T|)$ & $\mathcal{O}(L)$ & $\mathcal{O}(1)$ & non-native gates & circular & variational & circuit-based \\ \hline 
        \multicolumn{9}{c}{\rule{0pt}{11pt}Generative modeling} \\[3pt] 
        \hline
        QCBM\cite{Benedetti2019, Gili2022} & 2 & $\mathcal{O}(1)$ & $\mathcal{O}(N)$ & $\mathcal{O}(2^N)$ & native gates & linear & variational & shot-based \\ \hline 
        QNBM\cite{Gili2023} & 2 & $\mathcal{O}(1)$ & $\mathcal{O}(E)$ & $\mathcal{O}(2^{n_{\text{out}}})$ & classical control & all-to-all & variational & shot-based \\ \hline 
        \multicolumn{9}{c}{\rule{0pt}{11pt}Quantum neural networks} \\[3pt] 
        \hline
        QCNN\cite{Cong2019} & 2 & $\mathcal{O}(|T|)$ & $\mathcal{O}(N\lceil\log_{1/r}N\rceil)$ & $\mathcal{O}(1)$ & classical control & all-to-all & variational & circuit-based \\ \hline 
        QGNN\cite{Verdon2019} & 2 & $\mathcal{O}(|T|)$ & $\mathcal{O}(p)$ & $\mathcal{O}(1)$ & non-native gates & all-to-all & variational & circuit-based \\ \hline 
        \multicolumn{9}{c}{\rule{0pt}{11pt}Data analysis} \\[3pt] 
        \hline 
        NISQ-TDA\cite{Akhalwaya2022} & 2 & $\mathcal{O}(n_v)$ & $\mathcal{O}(V)$ & $\mathcal{O}(2^V)$ & classical control & all-to-all & non-variational & circuit-based \\ \hline
    \end{tabular}  
    \def\arraystretch{1}
    \begin{tabular}{rl|rl|rl}
        \footnotesize 
        $N$: & \rule{0pt}{9pt}number of qubits &\hspace{0.5cm} $t$: & number of Trotter time steps & $E$: & number of graph edges \\ 
        $\varepsilon$: & precision & $p$: & number of terms in Hamiltonian \hspace{0.5cm} & $n_{\text{out}}$: & number of output layer nodes \\ 
        $n$: & number of visible layers & $q$: & number of ansatz parameters & $r$: & reduction rate of pooling layers \\
        $m$: & number of hidden layers \hspace{0.5cm} & $|T|$: & cardinality of training set &\hspace{0.5cm} $V$: & number of graph vertices \\
        $n_p$: & number of particles & $L$ & number of re-uploading layers & $n_v$: & number of sampling vectors
    \end{tabular}
\end{table*}

This section features a practical assessment of selected quantum applications from the fields of quantum chemistry, quantum simulation, quantum machine learning, and data analysis in terms of their ARLs and extended classification labels (cf.\ subsection \ref{sec:QuantumUtility:ARL}).

A summary of all results can be found in Table \ref{tab:QUSurvey}. The following subsections contain details of the applications' individual investigations.

\subsubsection{Quantum chemistry and quantum simulation}

The idea to perform simulations of quantum systems on a quantum computer is as old as the field itself and is often attributed to Feynman\cite{Feynman1982}. Of particular interest is the study of electronic structure in the field of quantum chemistry, which has a wide range of applications within material science, drug discovery, and chemical engineering\cite{Tilly2022}. 

The variational quantum eigensolver (VQE)\cite{peruzzo2014variational} aims to efficiently estimate the ground state energy of a molecular system. Hence, it solves the variational optimization problem 
\begin{equation}
\min_{\theta} \braket{\Psi(\theta )|\hat{H}|\Psi(\theta )} = \min_{\theta} \braket{\Psi_0|\hat{U}^\dagger(\theta)\hat{H}\hat{U}(\theta)|\Psi_0}\,,
\end{equation} 
where $\hat{U}(\theta )\ket{\Psi_0}$ is assumed to be an accurate representation of the target ground state wave function and $\hat{H}$ denotes the molecular electronic Hamiltonian. There have been numerous advances in the field of VQE (see, e.g., \cite{Romero2019, Mizukami2020, Anselmetti2021, Gard2020, Wiersema2020, Barkoutsos2018, Wecker2015, Lee2019, Peruzzo2014, Kandala2017, Lang2021, Zhang2021, Ryabinkin2018, Yordanov2021, Zhang2022, Ryabinkin2020, Tang2021, Grimsley2019, Ostaszewski2021, Jordan1928, Bravyi2002, Seeley2012, Jiang2020, Shee2022, Cotler2020, Hadfield2021, Hadfield2020, Gokhale2019_2, Gokhale2019, Hamamura2020, Yen2021, Oumarou2022, Motta2021, Kivlichan2018, Huggins2021, Yen2020}). For a comprehensive review, refer to \cite{Fedorov2022, Tilly2022}.

Tabulated in Table \ref{tab:QUSurvey} are the respective best case scenarios for scalability, compilability, connectivity, robustness, and parallelizability. In terms of the VQE, sophisticated term grouping techniques such as explicit double factorization\cite{Huggins2021} or full rank optimization\cite{Yen2021} are claimed to reach reasonable energies for a linear scaling number of measurement circuits, $\mathcal{O}(N)$, w.r.t.\ the number of qubits $N$. All of these can be measured independently and therefore simultaneously on multiple quantum processing units (QPUs) such that VQEs possess a circuit-based parallelizability. Circuit ans\"atze of linear depth scaling have been developed (e.g., the k-UpCCGSD\cite{Lee2019} or the hardware-efficient ansatz (HEA)\cite{Kandala2017}). In case of the HEA, only native device gates and a linear qubit topology are required. The number of measurement shots is well-known to scale independently of the VQE system size with the squared inverse of the targeted accuracy of the expectation values.\cite{Tilly2022} 

In terms of its overall ARL, VQEs are currently based on level 3 since there are multiple resource estimations (see, e.g., references \cite{Tilly2022, Gonthier2022, Dalton2022}) that might indicate utility under special circumstances. At the same time, there have been no reports of classical (ideal or noisy) simulations that could prove this utility. It remains questionable if under the assumptions of linear scaling circuits, VQEs can still compete against the well-understood variety of classical quantum chemistry methods.

Recently, a quantum restricted Boltzmann machine (QRBM) approach was formulated by Xia and Kais to solve the electronic structure problem~\cite{Xia2018}. Their proposed QRBM is based on Carleo and Troyer's original work~\cite{Carleo2017} and can be identified as a direct competitor to the VQE framework. In contrast to VQEs, their QRBM involves a completely classical energy estimation. The central idea is to use the quantum computer as a sampling device that reproduces the optimal molecular wave function amplitudes in the measured probability distribution.

In their approach, each node of the classical RBM structure of a visible and a hidden neural network layer is mapped to individual qubits. Since all nodes of the visible layer need to be connected to all nodes of the hidden layer, the total circuit depth scales as $\mathcal{O}(nm)$ for $n$ visible and $m$ hidden layers. Furthermore, this high degree of connectivity demands an all-to-all qubit topology between both layers. 
Since they rely on specific post-measurement states, to perform the algorithm in an optimal fashion, classical control in the form of mid-circuit measurements is required. 

In contrast to VQEs, QRBMs have the advantage that they involve only a single measurement circuit but require $\mathcal{O}\binom{n}{n_p}$ measurement shots for $n_p$ particles in $n$ orbitals ($=$ visible layers) in the worst case scenario when every possible $n_p$-particle basis state needs to be sampled. Also, since there is only one circuit, multiple QPUs can only increase the total shot rate of that circuit. Overall, QRBMs reached ARL--2 with their entry publication~\cite{Xia2018} detailing a working proof of concept implementation. 

An alternative method to extract the ground state of a quantum many-body system is to evolve the Hamiltonian in imaginary time (under the Wick rotation $\tau = it$) \cite{Love2020}. Since the corresponding evolution operator $e^{-\hat{H}\tau}$ is non-unitary, imaginary time evolution may be a useful tool in simulating open quantum systems \cite{Kamakari2022}. In their recent publication \cite{McArdle2019}, McArdle et al. propose a variational quantum imaginary time evolution algorithm (VarQiTE). The key idea is in replacing the imaginary-time evolved state $\ket{\psi(\tau)}$ by the equivalent parametric state $\ket{\phi(\vec{\theta}(\tau))}$. Then, by applying the variational principle to the Wick-rotated Schr\"odinger equation, the authors are able to derive an analytic equation for the evolution of the parameters,
\begin{equation}
\label{VarQiTE.eq}
\sum_j A_{ij} \dot{\theta}_j = C_j,
\end{equation}  
where they provide simple quantum circuits to compute $A_{ij}$ and $C_j$. Given $t$ Trotter time steps, an $N$-qubit Hamiltonian with $p$ terms, and an ansatz $\phi(\vec{\theta}(\tau))$ with $q$ parameters, their algorithm requires $\mathcal{O}(tq(q+p))$ circuits to evaluate the matrix $A$ and vector $C$. Each circuit has depth $\mathcal{O}(q)$, and similar to VQE the number of shots scales independently of the system size. Advantageously, every circuit in each Trotter step can be implemented independently so the algorithm has circuit-based parallelizability. However, the algorithm as given requires all-to-all connectivity and uses non-native gates. The authors demonstrate the validity of their method on toy examples in quantum chemistry resulting in the ARL--2 classification. 

\subsubsection{Quantum machine learning}
In the context of computing, machine learning (ML) refers to a class of algorithms or computational paradigms where data is used to iteratively improve performance without explicit logic definition~\cite{Janiesch2020ML}. The tasks tackled with machine learning may include optimization, classification, and data generation. 

As a data-centric approach, the model must be able to capture the complexity of the training data. It can be difficult to extrapolate or \textit{generalize} to larger datasets using a simply trained model, and thus steps must be taken to ensure that the ML model is trained to be sufficiently general~\cite{Mohri2018Foundations}. 

There must be enough data points collected such that they can accurately capture the generalization of features within all possible input data - this can cause a tremendous increase in the amount of data points needed to train an ML model. Addressing the latter issue also requires powerful enough hardware to perform the training procedure in a reasonable amount of time, with a sufficiently fast clock speed and enough memory to store the parameters and intermediate calculations. With the advancement leaps being undertaken in the field of ML in recent years (see~\cite{Silver2016AlphaGo, Ramesh2022DALLE}), many of these issues have been addressed by concerted efforts to build larger and larger datasets (such as in~\cite{Deng2009ImageNet}) or training on supercomputing clusters (such as in~\cite{Smith2022Megatron}), though these advancements still require that one has access to these datasets and a sufficiently performant cluster, which can be a barrier to training and running ML models given a modest amount of time and resources.

These issues may be addressed using 
quantum machine learning (QML) models, which can be generalized using a smaller amount of training data than their classical counterparts\cite{Caro2022QMLFew}. Additionally, QML algorithms are capable of achieving equal or better prediction accuracy while making use of the same number of parameters, largely in part because of the accessible feature space of qubits being dictated by the traversable Hilbert space, which is exponential in the number of qubits, with significantly higher dimensionality than a feature space dictated by the same number of classical bits \cite{Mishra2021QMLReview}. This is not to mention the additional advantage that QML approaches have over classical ones in describing quantum data\cite{Huang2022QMLAdvantage}. As the field of machine learning (ML) reaches maturity, QML is still a nascent field with many opportunities for improvements.

\begin{center}
-- Binary and multi-class classification -- 
\end{center}

One of the central tasks tackled by ML algorithms is binary or multi-class classification. Given a $d$-dimensional data set $\Omega \subset \mathbb{R}^d$, which is assumed to be labeled by a map $m: \Omega \mapsto \{c_i\}$ to class labels $c_i$, the objective is to construct an optimal approximate map $\tilde{m}$ from only a given subset of all data points in $\Omega$. 

In the classical support vector machine (SVM) approach\cite{Vapnik1995}, a separating hyperplane is optimized in the data space to distinguish among differently labeled elements. This relies on the computation of inner products $\braket{\vec{x}_i|\vec{x}_j}$ in $\mathbb{R}^d$ for $\vec{x}_i,\vec{x}_j\in\Omega$. In general, such separating hyperplanes may be hard to find in the original (linear) feature space of the data set elements. Therefore, in the more general case, feature mappings $\phi$ are used to allow for complex non-linear representations via the kernel $\braket{\phi(\vec{x}_i)|\phi(\vec{x}_j)}$.

The potential benefits of a quantum kernel (QK) have been widely discussed\cite{Havlicek2019, Mensa2022}. The general idea is that relatively simple encoding circuits may reach non-trivial states in the underlying complex high-dimensional Hilbert space, which are hard to compute classically. In a straightforward QK approach~\cite{Schuld2019}, the quantum device can be used to calculate the inner products $\braket{0|\hat{U}(\vec{x}_i)^\dagger \hat{U}(\vec{x}_j)|0}$. The targeted inner product is then directly related to the probability $p$ to measure an all-zero state, i.e., $|\braket{0|\hat{U}(\vec{x}_i)^\dagger \hat{U}(\vec{x}_j)|0}|^2 = p(0\ldots 0)$. The resulting kernel matrix can then be used in a consecutive classical support vector machine to find the separating hyperplane.

In contrast to the QK method, which still uses a classical computer to estimate the separating hyperplane, the quantum variational classifier (QVC) approach~\cite{Schuld2019, Havlicek2019} aims to find a separating hyperplane directly on the quantum device by optimizing a parametrized training circuit. While the QK method relies on an accurate, near-perfect, quantum execution, the QVC approach may -- at least to some degree -- compensate noise in the training procedure. Each data point in the training set requires a separate encoding circuit such that the total number of circuits scales linearly with $|T|$. To improve the overall efficiency, all of these circuits may be measured in parallel on separate QPUs.

As a final classification algorithm in this assessment, consider the universal single qubit classification approach using data re-uploading~\cite{PerezSalinas2020}. Here, the central idea is to circumvent the no-cloning theorem~\cite{Park1970, Wooters1982, Dieks1982} by repeatedly \glqq{u}ploading\grqq{} data to the qubit register. This is achieved by interleaving fixed encoding and parametrized training gates. In the original publication~\cite{PerezSalinas2020}, both single-qubit and multi-qubit classification circuits with and without entanglement are compared, where the multi-qubit with entanglement case was found to be the most accurate. The number of required measurement shots depends only on the measurement of a single qubit's state tomography and therefore scales constantly with all problem-relevant system sizes. 

In conclusion, all three assessed classification algorithms were able to reach ARL--2 because beneficial (compared to scaled-down classical competitors) proof of concept simulations were reported~\cite{Schuld2019, Havlicek2019, PerezSalinas2020, Mensa2022}. Detailed resource analyses for specific quantum hardware -- required to progress to ARL--3 -- are still missing in the literature. One of the most problematic issues might be a correspondence between the number of measurement circuits and the cardinality of the training data set $T$, which may contain millions of data points in some problems. Therefore (and in light of recent findings~\cite{Caro2022QMLFew}), it is reasonable to focus the search for quantum utility of classification applications to problems where only very limited training data is available. 

\begin{center}
    -- Generative modeling -- 
\end{center}

Generative models are ML algorithms that extract unseen, defining features from an input data set and output new unseen data of similar type.
More precisely, a generative model tries to approximate an unknown probability distribution from only a subset of data samples from the target distribution. Unsupervised generative modeling tasks are claimed to be one of the first applications that may lead to a quantum utility (see, e.g., \cite{Benedetti2019_2, Zoufal2019, Tian2022, Gili2023}).

Both the Quantum Circuit and the Quantum Neuron Born machines (QCBM and QNBM, respectively)~\cite{Benedetti2019, Gili2022, Gili2023} are designed for generative modeling. The main theoretical concept behind both approaches is rather straightforward: Each bit string $x$ corresponding to state $\ket{x}$ is mapped to a specific (discrete) measurement outcome of the targeted probability distribution. A variational parametrized quantum circuit is then trained such that the measured bit string probabilities $|\braket{x|\psi}|^2$ (according to Born's rule — hence the name) match the target distribution. 

In case of the QCBM~\cite{Benedetti2019, Gili2022}, a standard hardware-efficient ansatz requiring only hardware-native gates and a linear qubit connectivity may be used. The latter scales linearly with the number of qubits, $N$, and can -- at least in principle -- learn probability distributions of $2^N$ outcomes. However, to sample such distributions also requires an exponentially scaling number of measurement shots, $\mathcal{O}(2^N)$.

Recently, Gili et al. introduced non-linear activation functions in the form of quantum neurons~\cite{Cao2017} into the QCBM framework~\cite{Gili2023}. Conceptually, their circuits resemble multi-layered neural network structures that can involve an arbitrary number of nodes per layer. Each node (input, hidden, and visible) is represented by an individual qubit, whereas the qubit's connectivity is enforced by the layer connection in the underlying network. Connected network nodes need to be connected in the corresponding qubit topology. To enforce non-linear activations, \glqq{r}erun until success\grqq{} type circuits involving the classical control feature of mid-circuit measurements are used.

Overall, both the QCBM and the QNBM approach are currently based on ARL--2 because beneficial proof of concept simulations comparing to scaled-down classical competitors have been reported~\cite{Benedetti2019, Gili2022, Gili2023}. Detailed resource estimations for a given problem and a given quantum hardware to progress to ARL--3 have not been reported. 

\begin{center}
    -- Quantum neural networks --
\end{center}

The quantum convolutional neural network (QCNN)~\cite{Cong2019} represents a class of quantum neural network, that introduces trainable (parametrized) convolutional layers, and conditional pooling layers, which reduce the number of active qubits by employing conditional mid-circuit measurements. Given an arbitrary input state $\rho_{\text{in}}$ encoded to $N$ qubits, a QCNN contains $\log_{\tfrac{1}{r}}(N)$ total layers (containing consecutive convolutional and pooling layers) for qubit reduction rate $r$ denoting the fraction of qubits eliminated by each pooling layer. 

In contrast to other QNNs, QCNNs are known to be efficiently trainable~\cite{Cong2019} and do not exhibit Barren plateaus~\cite{Pesah2021}. Assuming linear scaling sub-circuits for encoding, convolutional, and pooling layers, their total circuit depth scales as $\mathcal{O}(N\lceil\log_{1/r}N\rceil)$. The required number of measurement shots, on the other hand, scales independent of the problem size assuming that sufficient numbers of pooling layers are introduced to condense the circuits to the same number of finally measured qubits. For a sufficiently large number of pooling layers, the required qubit topology exceeds a 3D nearest neighbor connectivity, effectively requiring an all-to-all connectivity. With a supervised learning problem in mind, QCNNs involve $\mathcal{O}(|T|)$ different quantum circuits, which may be measured on multiple parallel QPUs simultaneously. 

Another approach to overcoming the difficulties of training QNNs is the quantum graph neural network formalism (QGNNs) as presented by Verdon et al. in their recent pre-print \cite{Verdon2019}. A similar approach has recently been tested on a neutral atom quantum computer by Albrecht et al.\cite{Albrecht2023} Many problems of interest are described by quadratic Hamiltonians which have a natural graph structure where each vertex represents a qubit, and each edge indicates a pairwise interaction. The goal of QGNNs is to obtain a learned low-dimensional representation of the feature graph describing the problem Hamiltonian. In the original publication\cite{Verdon2019}, the authors start with a general representation of a QGNN and go on to propose several specific ans\"atze by enforcing conditions on the training parameters that arise from the physics of the particular problem. For example, by enforcing invariance under node permutations the authors demonstrate a quantum graph convolutional neural network (QGCNN) which turns out to be a general form of the QAOA ansatz. Other problems considered by the authors include learning Hamiltonian dynamics, graph clustering and determining graph isomorphism. 

Similar to QCNNs, QGNNs require $\mathcal{O}(|T|)$ circuits to train the network where $|T|$ is the size of the training set, each of which can be executed in parallel on separate QPUs. For the more constrained ans\"atze, each circuit has depth scaling with the number of terms in the Hamiltonian $\mathcal{O}(p)$. In general, a quadratic Hamiltonian involves interactions between all pairs of qubits so the training circuits require all-to-all connectivity. The authors demonstrate four potential applications (including those mentioned above) of QGNNs via numerical implementation, therefore achieving the ARL--2 classification. 

\subsubsection{Data analysis}

Topological data analysis (TDA) is used to analyze complex high-dimensional data. It aims to reduce such data structures to a small set of local and global signature values associated with interpretable analytical value~\cite{Akhalwaya2022}. In essence, it involves the estimation of the persistent homology~\cite{Ghrist2008} of a given data graph, which contains the changing pattern of the graph's Betti numbers. TDA is claimed to have many fields of applications~\cite{Akhalwaya2022} ranging from astrophysics~\cite{Cole2018}, neuroscience~\cite{Chad2015}, genetics~\cite{Rabadan2020}, and the analysis of deep neural networks~\cite{Naitzat2020}.

On classical computers, TDA can be a task of tremendous difficulty\cite{Lloyd2016}. Fault-tolerant quantum algorithms have been developed that may one day reach an exponential advantage over its classical competitors \cite{Lloyd2016} on real hardware. Recently, Akhalwaya et al.\ published a novel approach to TDA in the NISQ era of quantum computing called NISQ-TDA \cite{Akhalwaya2022}. In their approach, the TDA problem is reframed to a stochastic matrix rank estimation using the Chebyshev method \cite{Ubaru2016, Ubaru2017}. The full quantum algorithm requires $N$ qubits for $N$ data points and involves (i) the generation of a random initial state of zero mean, (ii) the application of a counting algorithm using $\log_2N$ ancilla qubits, (iii) the application and mid-circuit measurement of CCNOT gates onto ancilla qubits controlled by all connected (within the applied distance threshold) data points, and (iv) the application of a Jordan-Wigner mapped boundary operator. The total circuit depth is claimed to scale linearly with the number of data graph vertices, $V$, for a fixed polynomial degree on quantum devices that support mid-circuit measurements~\cite{Akhalwaya2022}. Since their final measurement involves to estimate the norm of their projected quantum states, the number of required measurement shots scales exponentially with the number of analyzed data graph vertices. Explicit resource estimations in terms of, e.g., runtimes for a specific quantum device running a classically demanding TDA problem were not reported. Therefore, NISQ-TDA is classified as ARL--2 but shows promising features to progress to ARL--3. 

\section{Conclusion}
In this paper, the general concept of a practical quantum advantage -- coined by the term \textit{quantum utility}
(cf.\ Def.\ \ref{def:QU}) -- and its assessment in terms of \textit{application readiness levels} (ARLs, cf.\ 
Figures \ref{fig:ARL_Roadmap} and \ref{fig:ARL_Milestones}) was introduced. 

In essence, quantum utility represents 
a milestone that is coupled to both the practical application and the quantum device running the latter. 
To assess the 
maturity of a given application in the pursuit of quantum utility for a fixed quantum device, subsection 
\ref{sec:QuantumUtility:ARL} introduced ARLs. These were deeply inspired by NASA's technology readiness levels (TRLs)
\cite{ISO_TRL_2013} defining a five level scheme highlighting an application's progress towards quantum utility via 
(ARL--1) conceptualization, (ARL--2) proof of concept, (ARL--3) proof of scalability, (ARL--4) utility in simulation, 
and finally (ARL--5) quantum utility. 
Furthermore, extended classification labels (cf.\ Table \ref{extended_labels.tab}) were provided to ease
utility investigations in a compact and informative fashion by detailing a quantum application's requirements in 
terms of scalability, gate compilability, qubit connectivity, robustness to noise, and parallelizability 
to multiple quantum units. 

To demonstrate the ARL classification scheme, section \ref{sec:practicalassessment} includes a survey of different 
quantum applications from the fields of quantum chemistry, quantum simulation, quantum machine learning, and 
data analysis. Each application was assessed in terms of both their current ARLs and their extended 
classification labels in the respective best case scenarios. All results were collected in 
Table \ref{tab:QUSurvey} and briefly discussed in the consecutive subsections. 

As of yet, VQEs are the only application from the presented survey that were able to reach ARL--3. This is because 
detailed resource estimations, e.g., full algorithm QPU timings, are seldom reported in the literature -- 
especially in the case of similar device comparisons as targeted by quantum utility.   
However, it is reasonable to assume that many of the displayed applications can and will easily progress to ARL--3 or 
higher given their promising scaling and algorithm features. 

This paper may be understood as a \glqq{c}all to arms\grqq{} for the quantum industry, researchers, 
algorithm developers or anyone interested, to pursue, think about and aim for quantum utility in their developments.
Especially mobile or edge-based applications may provide a much more tangible target when tackled by edge-deployable 
quantum hardware of appropriate size, weight, and cost. We strongly believe such applications to be 
among the first to demonstrate a genuine quantum advantage, paving the way to truly ubiquitous quantum computing.

\providecommand{\noopsort}[1]{}\providecommand{\singleletter}[1]{#1}%

\end{document}